\documentclass[aps,prl,twocolumn,showpacs,floatfix]{revtex4}
\usepackage{amsmath}
\usepackage{amssymb}
\usepackage{graphicx}
\usepackage{subfigure}
\usepackage{dcolumn}
\usepackage{bm}

\begin{document}

\title{How to Define Variation of Physical Properties Normal to an Undulating One-Dimensional Object}
\author{Hsiao-Ping Hsu} \author{Kurt Binder} \author{Wolfgang Paul \email{Wolfgang.Paul@Uni-Mainz.De}}
\affiliation{Institute of Physics, Johannes-Gutenberg University,
  55099 Mainz, Germany}

\date{\today}

\begin{abstract}
One-dimensional flexible objects are abundant in physics, from
polymers to vortex lines to defect lines and many more. These objects
structure their environment and it is natural to assume that the
influence these objects exert on their environment depends on the
distance from the line-object. But how should this be defined? We
argue here that there is an intrinsic length
scale along the undulating line that is a measure of its 
``stiffness" (i.e., orientational persistence),
which yields a natural way of defining the variation of physical
properties normal to the undulating line. We exemplify how this normal
variation can be determined from a computer simulation 
for the case of a so-called bottle-brush polymer, where side
chains are grafted onto a flexible backbone.
\end{abstract}
\maketitle

One dimensional objects are characteristic for the physics of quite
diverse phenomena. Most prominent may be linear polymers \cite{polymref1,polymref2}, be they
synthetic ones or biopolymers \cite{bioref1}, and vortex matter in high $T_c$
superconductors \cite{blatter,besseling}. But line objects also
determine the characteristics of the structure of (poly-)crystals in the form of defect
lines \cite{crystal-line} and they were argued to determine the
structure of metallic glasses \cite{nelson} in the form of
disclination lines. This list, of course, makes no claim to
completeness. In all cases, however, physical properties around the
line objects will change depending on distance to the object. For
polyelectrolyte chains like, e.g., DNA, one is for instance interested
in the distribution of counterions around the chain \cite{manning} and the
phenomenon of counterion condensation on the (linear) chain. In the case of vortex lines, the
order parameter distribution around these lines is important
\cite{vortex-ref}, and there exists also a rich phenomenology of
interactions between vortex matter and the defect structure of the
underlying lattice \cite{verdene}. It is also well known that dislocation
lines in metals control their plastic deformation. 
Defect lines in ordered materials and the strain fields around them are directly observable in colloidal
crystals and nematic materials \cite{allen,smalyukh,callan} and are of
importance for instance for the performance of photonic crystals
\cite{snow}. In all these cases, physical quantities can be expected to change in space
according to the distance of a given point from the line object. Such
a normal or radial distance is clearly defined for a straight line,
and sometimes one uses models assuming infinite rigidity to look at
radial variation, like for example the cell model of polyelectrolytes
\cite{cell-model}, or chain molecules tethered to a straight line as a model for a
molecular bottlebrush \cite{hsu}, etc. However,the usefulness of all these models
has been completely uncertain. In reality, however, these line objects are not
infinitely rigid and thus they undulate due to thermal
fluctuations. This undulation leads to an interaction between
different parts of the lines through the 
structuring effect they have on their environment. The question of how to
define a radial distance from the line object in this case
naturally arises when one studies such physical systems with computer
simulations, but also for experiments \cite{pires}, when the structure of a system
can be observed for instance microscopically or tomographically.

From the differential geometry of space curves we have an easy answer
to this problem locally. Just calculate the two normal vectors (in
3d) to the line. And globally, when the line object is curled up to 
a globule or coil, 
there is no natural normal direction. While the latter is true also
when we take into account the granularity of matter, the local
definition seizes to be helpful on the atomic scale. Here we will find
preferred angles between consecutive segments of the line given either
by the underlying chemical nature of the object like in polymers or by
an underlying lattice structure like in vortex matter or defect
lines. The natural length scale for a definition of normal variation
will therefore lie in between this local - and very system specific -
and the global - in many cases isotropic - scale. We will discuss in
the following how an optimal choice for this intermediate length scale
can be identified both in simulations and in experiments providing
real space structural information. The procedure will be completely
general, however, we will exemplify it for the case of a so-called
bottle brush polymer \cite{hsu}. 

\begin{figure}[htb]
\vspace*{5mm}
\begin{center}
\includegraphics[width=0.9\columnwidth]{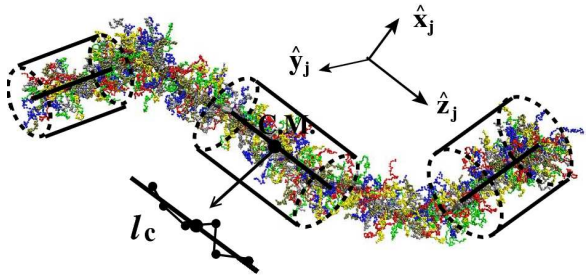}
\caption{Snapshot of a bottle brush polymer simulated using the bond
  fluctuation model. The backbone consists of $N_b=387$ repeat units
  and onto each of these a side chain of length $N=48$ is grafted. One
  segment of the backbone consisting of $l_c$ bonds is plotted next to
the chain. Such segments define a local coordinate system also
indicated in the figure and cylinders surrounding the simulated chain
as shown (see text).} 
\label{fig1}
\end{center}
\end{figure}
Our line object will therefore locally be made out of straight segments - we
call them bonds from now on - making up 
the linear backbone of a bottle brush polymer shown in
Fig.~\ref{fig1}. We regard a segment of the backbone of this bottle
brush made out of $l_c$ consecutive bonds. For this segment we can
define an equivalent cylinder axis as follows: each unit
bond vector $\hat{e}_i$ can be seen as the normal to a plane, in which we
define cylinder coordinates. The average orientation of these planes
is given as
\begin{equation}
\hat{n} = \frac{1}{l_c} \sum_{i=1}^{l_c} \hat{e}_i\; .
\end{equation}
This definition of $\hat{n}$ is similar to how one would define the
normal to an undulating surface. When we assume a triangulation of the
surface, each triangle defines a unique normal vector, and the average
over these normal vectors for adjacent triangles defines the surface 
normal on a coarse-grained scale.
Using $\hat{n}$ we can define an equivalent straight cylinder for our
undulating line on the coarse-grained scale $l_c$. 
The z-axis is given by $\hat{n}$, it goes through the center of mass
(c.m.) of the backbone segment made of $l_c+1$ monomers and has a length
given by the end-to-end distance $R_e^c =
\|\vec{r}_{n+l_c}-\vec{r}_n\|$ of the backbone segment between monomers
$n$ and $n+l_c$. In the 
normal plane to this axis we employ cylinder coordinates to define
the radial distance of a given point to the axis of the cylinder. 
Each point in space is considered to count for the normal variation
within this cylinder segment, when its z-coordinate in the local
coordinate frame lies between $z=0$ and $z=R_e^c$. By construction,
each point in space may lie in more than one of these cylindrical
segments, in which case it is, however, only counted at the smallest
distance to these segments 
(when degeneracy occurs where a monomer has the same distance
to the axis in two cylinder segments, it should be counted for both
segments with weight $1/2$). Evaluating whichever scalar or tensorial
physical property is characteristic for the considered system at these points, one
can in this way map out a variation of this property normal to the
undulating line for which one performed this construction. 

In our case of bottle brush polymers we study the normal variation of
the density of side chain monomers with respect to the backbone of the
polymer. This density variation is something one typically tries to
extract from scattering experiments \cite{schmidt,rathgeber,fenz,hsu2}. Due to
unavoidable approximations in the analysis of the scattering data, there has
been, however, a controversy in the literature \cite{schmidt,rathgeber} about
the form of this spatial variation. To resolve this controversy we apply our
procedure of defining what is meant by variation normal to the backbone of the
bottle-brush polymer.
Our construction ensures that each monomer is contributing not
more than once to the determination of the radial density
profile; however near the backbone chain ends some monomers
may not be counted at all
(when their
z-coordinates do not fall into the allowed range for any local
cylindrical coordinate system). This is an example of chain end effects
occurring whenever the undulating line has only a finite length.
\begin{figure}[htb]
\vspace*{5mm}
\begin{center}
\includegraphics[width=0.7\columnwidth,angle=-90]{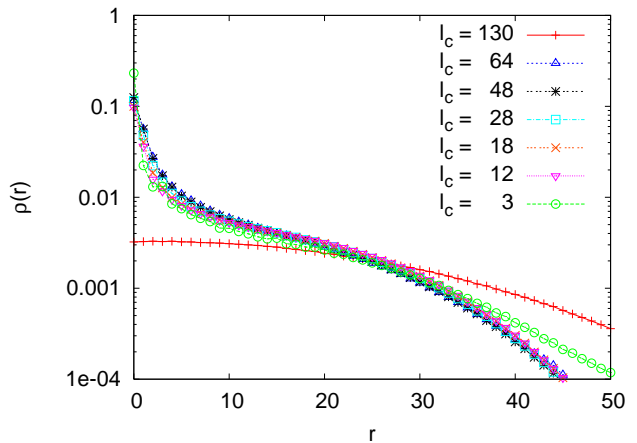}
\caption{Variation of the density of side chain monomers normal to the
backbone of the bottle brush for $N_b=131$ and $N=48$. Results for
several choices of the coarse-graining length $l_c$ along the backbone
are shown. All lengths are measured in units of the lattice spacing of
the underlying simple cubic lattice.} 
\label{fig2}
\end{center}
\end{figure}
The normal variation defined in this way
depends on the coarse-graining length scale $l_c$ along the undulating
line. Figure \ref{fig2} shows the dependence
of the normal density on this length scale for a backbone of length
$N_b=131$ and side chain length $N=48$. The data are
for a Monte Carlo simulation of the bond-fluctuation lattice model
\cite{bflref}. Clearly, the
form of the curves for very small ($l_c=3$) and very large ($l_c=130$)
values of the coarse-graining length significantly differs from the
behavior for intermediate scales. The behavior for small $l_c$
gives comparable results to an approach taken in \cite{holm} to measure
the counterion concentration around a polyelectrolyte chain. We now want to argue that at the
intermediate scales there exists an (almost sharply defined) optimal
length scale for coarse-graining.

\begin{figure}[htb]
\vspace*{5mm}
\begin{center}
\includegraphics[width=0.7\columnwidth,angle=-90]{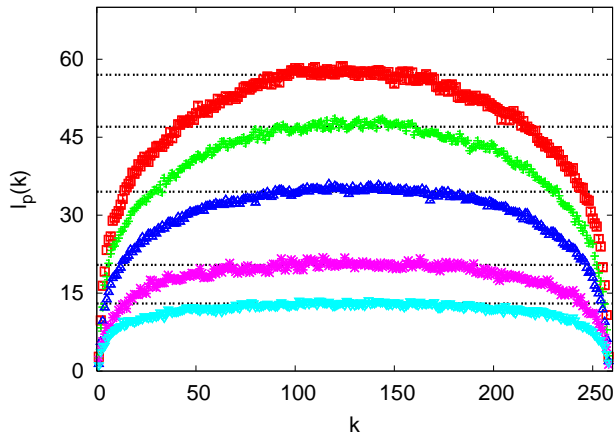}
\caption{Projection of the end-to-end vector of the backbone of the
  bottle brush onto the local bond orientation as a function of
  position of the bond along the backbone. The backbone length is
  $N_b=259$ and the side chain lengths are $N=6, 12, 24, 36$ and $48$ from
bottom to top. Dotted lines indicate values $l_p^{(1)}$ of the
persistence length.} 
\label{fig3}
\end{center}
\end{figure}
This length scale has to be connected with the intrinsic stiffness of
the undulating line object. In polymer physics this is called the
persistence length of the polymer chain (here the backbone of the
bottle brush). The persistence length is known to be the natural scale for a
coarse-graining of intramolecular behavior, i.e., variation as a function of
chemical distance along the chain. We suggest here, that it is also a natural
scale for construction of a coarse grained description of property variation
as a function of spatial distance to the chain. There are several
prescriptions of how to obtain the
persistence length from an ensemble of configurations of a
chain. One definition uses the projection of the end-to-end vector of
the chain onto the unit bond vector of bond $k$, measured in terms of the
average length of this bond
\begin{equation}
l_p(k) = \langle \vec{b}_k \cdot \vec{R}_e /{| \vec{b}_k|^2}\rangle\; .
\end{equation}
In a plot of $l_p(k)$ as a function of position $k$ along the chain
as shown in Fig.~\ref{fig3} one can clearly identify a plateau
regime. The height of this plateau defines the persistence length
$l_p^{(1)}$. Another definition employs the bond vector orientational
correlation function $\langle \cos \theta(s)\rangle = \langle
\hat{e_i}\cdot\hat{e}_{i+s} \rangle$ where the angular brackets
include an average over $i$.
For idealized polymer models this function decays exponentially as
$a_o\exp\{-s/l_p^{(2)}\}$ (the amplitude prefactor is necessary for
discrete models for which the exponential does not extrapolate to one
for $s\to 0$). In reality, this function generally is not a
single exponential, and we propose to define the persistence length
for arbitrary undulating lines as 
\begin{equation}
   l_p^{(2)} = \frac {1}{a(N_b-1)} \sum_{s=1}^{N_b-1} <\cos \theta(s)> \;.
\end{equation} 
If $l_p^{(2)}>>1$ and the decay of $<\cos \theta(s)>$ with $s$ is a single
exponential, one can for $N_b \rightarrow \infty$ transform the sum to an
integral and obtains again the standard definition of $l_p^{(2)}$.
It must be noted, however, that for real polymer chains (which
have excluded volume interactions) one finds power law decays rather than
exponential decay for $N_b \rightarrow \infty$\cite{24,25,26}.
For isolated self-avoiding walks (good solvent conditions)
$< \cos \theta(s)> \; \propto \; s^{2\nu-2}$ and then
$l_p^{(2)} \propto l_p^{(1)} \propto N_b^{2\nu-1}$\cite{24}.
In dense melts $< \cos \theta(s)> \; \propto \; s^{-3/2}$~\cite{25},
$l_p^{(2)}$ \{Eq. (3)\} then is dominated by the behavior for small $s$,
and does not depend on $N_b$ for lager $N_b$.
Thus the use of Eq.(3) requires care.
Both estimates of the persistence
length agree within $10\%$ with each other and show an increase by a factor
of about $4$ (see Fig.~\ref{fig3}) when we increase the side chain length from $N=6$ to
$N=48$. This is in excellent agreement with experiment
\cite{schmidt-persistence} and agrees with simulations in
\cite{rath-pak} but is a stronger variation than found in
\cite{arun}. For the present purposes the important
point is that the stiffness length scale $l_p$ (and hence the
coarse-graining length scale $l_c$ in Fig.~\ref{fig1}) can be varied
over a wide range, which allows us to assess the validity of our
general concepts. 

\begin{figure}[htb]
\vspace*{5mm}
\begin{center}
\includegraphics[width=0.7\columnwidth,angle=-90]{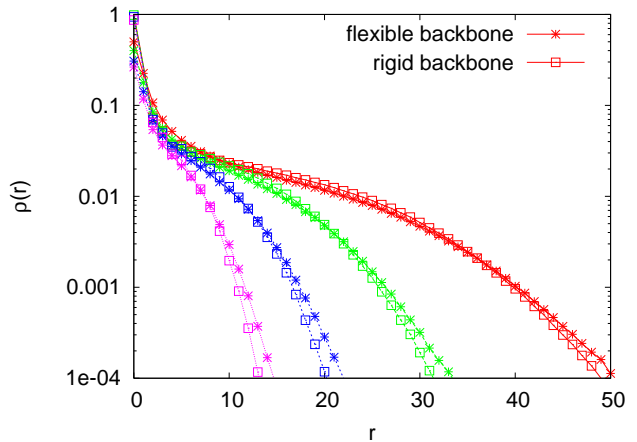}
\caption{Density variation of side chain monomers normal to an
  undulating backbone of length $N_b=131$ for the choice $l_c=l_p$ and for the rigid
  backbone. The four choices of side chain length are $N=6, 12, 24$
  and $48$ from left to right.} 
\label{fig4}
\end{center}
\end{figure}
When we now take the persistence length as the natural scale for the
definition of the coarse-graining length in Fig.~\ref{fig1}, i.e. $l_c =
l_p$, we find that the normal variation determined in this way is in
excellent agreement with what one would determine for a completely
rigid backbone. In Fig.~\ref{fig4} we compare the radial density
variation for a backbone length $N_b=131$ and four different side
chain lengths $N=6,12,24,48$ obtained using $l_c=l_p$ to the radial
density variation around a rigid backbone
\footnote{For the rigid backbone case, all backbone bond vectors were chosen as $(003)$
and motions of side chain monomers were performed as in the flexible case.
Averaging of $\rho(r)$ was done for the full backbone length.}.
Obviously, our definition
of normal variation with respect to the undulating backbone for this
choice of coarse-graining length nicely agrees with the variation
around the rigid backbone at small to intermediate distances. In this way,we
justify the often used rigid backbone model for densely grafted
bottle-brushes. For large distances, the density measured for a flexible backbone has to
be larger than the one measured for the rigid backbone, as the former
contains contributions from remote segments of the backbone which can
bend back on itself. From our experience, we can vary $l_c$ by
about $20$~\% around the choice $l_c=l_p$ without much affecting the
agreement observed in Fig.~\ref{fig4}. Therefore, the persistence
length provides the natural length scale for the coarse-graining
depicted in Fig.~\ref{fig1}. Reinterpreting Fig.~\ref{fig4} from a
different perspective, we can say that an analysis of the variation
of physical properties normal to an infinitely stiff line-object gives
a good approximation to the behavior for real, thermally fluctuating
lines on the scale of the persistence length. 

To conclude, we have outlined a constructive procedure how to define
and measure normal variation of physical properties with respect to an
undulating line. This problem is relevant for a broad range of
physical situations ranging from linear polymers, to vortex matter to
defect lines in crystals. The procedure is applicable for simulations as well
as for experiments which provide real space structural information. We
explained how a local cylindrical coordinate system can be defined on
a coarse-grained scale along the line, in which the normal variation
with respect to the undulating line object can then be
determined. The optimal length scale
for the introduction of cylindrical 
reference frames turned out to be the persistence length of the fluctuating line
object. We showed for the example of the density variation around the
backbone of a bottle brush polymer that a normal variation around a
flexible backbone determined for this length scale reproduces
the normal variation around a rigid backbone on short and intermediate
scales, where they should ideally be identical.  
This example is important, since such stimuli-responsive polymers may find
applications as sensors and actuators \cite{li}, and for this purpose, as well
as for asserting biomolecular functions of bottle-brushes such as
aggrecane in mammalian joints \cite{klein}, a deeper understanding of bottle-brush
properties is required.

For a reliable interpretation
of experimental data on the variation of physical properties around an
undulating line object one has to go beyond the 
approximation of infinite rigidity. This manuscript suggests a
controlled way to do this and to assess to what extent theoretical
descriptions assuming infinite rigidity are valid.
Besides the realm of synthetic and bio-polymers, we envisage potential
applications of our concepts to problems such as the radial
distribution of vacancies, solute atoms, etc., around lines in
dislocation networks in crystals and liquid crystals. Also,
heterogenous nucleation along line defects will be affected by their
local curvature and it will be interesting to elucidate the
differences to the case of completely rigid defect lines.   

{\bf Acknowledgment:} We acknowledge funding through the German
Science Foundation through the collaborative research project SFB 625,
sub-project A3. We are grateful to the J\"ulich Supercomputer Center
for computer time on the JUMP computer through project HMZ03, and to
the European network of excellence SoftComp for computer time on the
SoftComp computer cluster.


\newpage

\begin{thebibliography}{00}
\bibitem{polymref1} A. Yu Grosberg, A. R. Khokhlov, {Statistical
  Physics of Macromolecules}, (AIP Press, Woodbury, 2002).
\bibitem{polymref2} T. A. Witten, Rev. Mod. Phys. {\bf 70}, 1531
  (1998). 
\bibitem{bioref1}Zwolak M. and Di Ventra M., Rev. Mod. Phys. {\bf 80},
  141 (2008).
\bibitem{blatter} G. Blatter et al. Rev. Mod. Phys. {\bf 66}, 1125
  (1994). 
\bibitem{besseling} R. Besseling, N. Kokubo and P. H. Kes,
  Phys. Rev. Lett. {\bf 91}, 177002 (2003).
\bibitem{crystal-line} M. Kleman, J. Friedel, Rev. Mod. Phys. {\bf
  80}, 61 (2008).
\bibitem{nelson} S. Sachdev and D. R. Nelson, Phys. Rev. Lett. {\bf
  53}, 1947 (1984).
\bibitem{manning}G. S. Manning, J. Chem. Phys. {\bf 51}, 924 (1969).
\bibitem{vortex-ref}T. Nattermann and S. Scheidl, Advances Physics
  {\bf 49}, 607 (2000).
\bibitem{verdene} T. Verdene et al., Phys. Rev. Lett. {\bf 101},
  157003 (2008).
\bibitem{allen} D. Andrienko and M. P. Allen, Phys. Rev. E {\bf 61},
  504 (2000). 
\bibitem{smalyukh} I. I. Smalyukh et al., Mol. Cryst. Liq. Cryst. {\bf
  450}, 79[279] (2006).
\bibitem{callan}A. C. Callan-Jones et al., Phys. Rev. E {\bf 74},
  061701 (2006).
\bibitem{snow} B. D. Snow et al., Mol. Cryst. Liq. Cryst. {\bf 502},
  178 (2009).
\bibitem{cell-model}M. Deserno, C. Holm and S. May, Macromolecules
  {\bf 33}, 199 (2000).
\bibitem{hsu} H.-P. Hsu, W. Paul and K. Binder, Macromol. Theory
  Simul. {\bf 16}, 660 (2007).
\bibitem{pires} D. Pires, J. B. Fleury and Y. Galerne,
  Phys. Rev. Lett. {\bf 98}, 247801 (2007).
\bibitem{schmidt}B. Zhang et al., Macromolecules {\bf 39}, 8440
  (2006). 
\bibitem{rathgeber}S. Rathgeber et al., J. Chem. Phys. {\bf 122},
  124904 (2005).
\bibitem{fenz} L. Fenz et al., Eur. Phys. J. E {\bf 23}, 237 (2007). 
\bibitem{hsu2} H.-P. Hsu, W. Paul and K. Binder, J. Chem. Phys. {\bf
  129}, 204904 (2008).
\bibitem{bflref}K. Binder and W. Paul, Macromolecules
  {\bf 41}, 4537 (2008).
\bibitem{holm} H. J. Limbach and C. Holm, J. Phys. Chem. B {\bf 107},
  8041 (2003).
\bibitem{24} L. Sch\"afer and K. Elsner, Eur. Phys. J. E {\bf 13}, 225 (2004).
\bibitem{25} J. P. Wittmer et al., Phys. Rev. Lett. {\bf 93}, 147801 (2004);
Phys. Rev. E {\bf 76}, 011803 (2007).
\bibitem{26} D. Shirvanyants et al., Macromolecules {\bf 41}, 1475 (2008).
\bibitem{schmidt-persistence} N. Gunari, M. Schmidt and A. Janshoff,
  Macromolecules {\bf 39}, 2219 (2006).
\bibitem{rath-pak} S. Rathgeber et al., Polymer {\bf 47}, 7318 (2006).
\bibitem{arun} A. Yethiraj, J. Chem. Phys. {\bf 125}, 204901 (2006).
\bibitem{li} C. Li et al., Angew. Chem. Int. Ed. {\bf 43} ,1101 (2004).
\bibitem{klein}J. Klein, Science {\bf 323}, 47 (2009).
\end{thebibliography}
\end{document}